\documentclass[aps,prd, preprintnumbers,groupedaddress,nofootinbib,twocolumn]{revtex4}

\pdfoutput=1
\usepackage{graphicx}
\usepackage{latexsym}
\usepackage{amsfonts}
\usepackage{amssymb}
\usepackage{amsmath}
\usepackage{slashed}
\usepackage{feynmp}
\usepackage{hyperref}

\begin{document}

\title{Asymmetric Dark Matter and Effective Operators}

\author{Matthew R.~Buckley$^{1}$}
\affiliation{$^1$Center for Particle Astrophysics, Fermi National Accelerator Laboratory, Batavia, IL 60510}
\preprint{FERMILAB-PUB-11-168-A}

\begin{abstract}
In order to annihilate in the early Universe to levels well below the measured dark matter density, asymmetric dark matter must possess large couplings to the Standard Model. In this paper, we consider effective operators which allow asymmetric dark matter to annihilate into quarks. In addition to a bound from requiring sufficient annihilation, the energy scale of such operators can be constrained by limits from direct detection and monojet searches at colliders. We show that the allowed parameter space for these operators is highly constrained, leading to non-trivial requirements that any model of asymmetric dark matter must satisfy.
\end{abstract}
\maketitle

Despite decades of experimental effort, remarkably little is known about the nature of dark matter. For many years, the leading theoretical class of candidates for dark matter has been a Weakly Interacting Massive Particle (WIMP). The success of this paradigm is due in large part to the surprising fact that a thermal relic with a weak-scale mass and interaction strength will have the correct dark matter abundance. However, it should be noticed that in most phenomenologically viable models, some level of fine-tuning is necessary, weakening the motivations behind the `WIMP miracle' (see, for example Ref.~\cite{Feng:2009qf}).

Recently, a proposal for an alternative origin of dark matter has gained in prominence: that of asymmetric dark matter (ADM) \cite{Kribs:2009fy,Cohen:2009fz,An:2009vq,Cohen:2010kn,Kaplan:2009ag,Buckley:2010ui,Davoudiasl:2010am,Belyaev:2010vn,Graesser:2011wi,Haba:2010bm,Shelton:2010ta,Blennow:2010qf,Frandsen:2011kx} (for earlier works along similar lines, see Refs.~\cite{Agashe:2004bm,Banks:2006xr,Cosme:2005sb,Farrar:2005zd,Hooper:2004dc,Kaplan:1991ah,Kitano:2004sv,Kitano:2008tk,Suematsu:2005kp,Thomas:1995ze,Tytgat:2006wy}). In this class of models, the coincidence of energy densities of baryons and dark matter (which differ only by a factor of $\sim 6$) is taken as the driving motivation. This leads to the conclusion that dark matter, like baryons, should be composed of a particle $\chi$ with a quantum number $X$ which is conserved at low energies and generated through some $X$-violating process, rather than consisting of a thermal bath of $\chi/\bar{\chi}$ particles with the $X$ number of the Universe equal to zero. The similarity of the baryon and dark matter densities suggests that the $X$-violating process should somehow be connected to $B$ or $L$ number violating processes that must have occurred in early Universe baryogenesis. The larger density of dark matter can then be explained either through a dark matter mass $m_\chi$ of the order $4-10$~GeV (such models include darkogenesis \cite{Shelton:2010ta} and hylogenesis \cite{Davoudiasl:2010am}), or by a much heavier dark matter mass (weak scale or above) combined with a mass suppression during the era of $X-B$ transfer (Xogenesis \cite{Buckley:2010ui}).

The large number of proposed ADM models differ wildly in their explanation of the origin of the $X$ asymmetry, the mechanism of transfer of asymmetry from the dark to the visible sectors, and the required mass $m_\chi$. However, there is one universal requirement that every model must meet: the thermal relic density of $\chi/\bar{\chi}$ (the symmetric component of dark matter) must be compose only a small fraction of dark matter's total contribution to the Universe's energy budget.\footnote{While it is certainly possible for both symmetric and asymmetric components to contribute significantly, this requires multiple coincidences in the operators responsible for both transfer and annihilation. Such a model may be found in Ref.~\cite{Graesser:2011wi}. While in this paper we shall not consider this possibility in more depth, we include results applicable to pure symmetric dark matter, allowing the reader to interpolate the results for a mixed scenario.}

Since the contribution to the matter density of the symmetric component is much less than $\Omega_{\rm DM}$, the thermal cross section in the early Universe must be significantly larger than that usually assumed for a WIMP. Thus, in ADM either there must be large couplings between the dark matter and some visible sector particles, or additional very light states in the dark sector into which the dark matter can annihilate without over-closing the Universe. In the former scenario, the required large interactions with the Standard Model may result in direct detection cross sections that can be probed by current experiments.

In this paper, we consider effective operators between two dark matter particles and two quarks. The effective operator formalism allows us to remain agnostic as to the particle content at high energy scales, by considering only operators that respect Standard Model gauge invariance after electro-weak symmetry breaking and couple the dark matter directly to the Standard Model fields. In order to include the low-energy effects of any unknown high-mass particles, we add operators to the Lagrangian that are of dimension greater than four. Such operators must be suppressed by an energy scale $\Lambda$, which is roughly equivalent to the mass of the mediating particle over the coupling at the high scale. 

A familiar example of effective operators is the four-fermion interaction, which accounts for the weak interaction at scales much less than the mass of the $W$ and $Z$ bosons. In this case, the dimension six operator is suppressed by the Fermi constant, which in the language of this paper would be expressed as $G_F= \Lambda^{-2}$. In this particular example $\Lambda$ would be defined as $2^{5/4}m_W/g$, where $m_W$ is the $W$ boson mass, and $g$ is the weak coupling constant.

Such effective operators preserve both $X$ and $B$, and so are not related to the origin of dark matter or baryons, however they are necessary components for any successful ADM model. Assuming that the symmetric component of dark matter makes up less than $10\%$ of the total $\Omega_{\rm DM}$, we place upper bounds on the suppression scale $\Lambda$ for each operator for both complex scalar and Dirac fermion dark matter.\footnote{Majorana fermions and real scalars possess no conserved global current $X$, and so are not good candidates for ADM. Small Majorana masses -- leading to $\chi-\bar{\chi}$ oscillations on cosmological timescales -- are not ruled out in ADM models, but can be ignored for the purpose of this paper.}  Comparison with the predicted direct detection cross section with the current experimental bounds can then used to place lower bounds on $\Lambda$ for many of the operators. Monojet plus missing energy ($\slashed{E}_T$) searches at the Tevatron, which would arise from pair production of dark matter plus a jet (used for the event trigger) can also place lower limits on the suppression scale \cite{Bai:2010ys,Goodman:2010ly,Goodman:2010zr}.

As we shall show, these bounds place severe restrictions on the allowed range of the scale $\Lambda$; in fact, they completely exclude the entire parameter space for several classes of operators. From this, we can greatly constrain the possible interactions for any asymmetric dark matter model.  In using the effective operator formalism, this paper has similarities to the work of Refs.~\cite{Bai:2010ys,Goodman:2010ly,Goodman:2010zr,Fox:2011tg}, which consider the bounds on effective operators for symmetric dark matter. As we will show, the application of these bounds to the asymmetric dark matter leads to some very interesting conclusions: namely that (outside some tightly constrained regions of parameter space) a successful model of asymmetric dark matter must contain new light states, leptophilic couplings, or new confining gauge interactions. These conclusions should be taken into account when considering motivations for asymmetric dark matter model-building.

There are, of course, two major assumptions underlying this approach which deserve to be stressed at this point. First, that the same operator that over-annihilates dark matter in the early Universe is active today, and second that the annihilation operator allows for couplings of dark matter to quarks. The latter assumption allows the operators to be bounded by results from direct detection and hadronic collider experiments, though leptophilic dark matter can be probed by LEP searches \cite{Fox:2011tg} instead. Dark matter which annihilates into some new light state of the dark sector is much more difficult to probe, though by the assumption of ADM such states must be light enough not to dominate the matter density, while also evading BBN constraints on relativistic degrees of freedom. 

The assumption that a single operator is responsible for both direct detection and over-annihilation is primarily made for simplicity: the derived bounds on operators would not apply to scenarios with (for example) composite dark matter \cite{Nussinov:1985xr,Chivukula:1989qb,Bagnasco:vn,Khlopov:2008ly,Frandsen:2011kx,Gudnason:zr} or dark atoms \cite{Kaplan:2009kx}. In both cases the present-day direct detection cross sections are suppressed by form factors (though the collider bounds would be unaffected). These assumptions are fairly strong, but -- as will be demonstrated -- the operators considered in this paper are highly constrained. Therefore, should future direct detection and collider bounds completely rule out the operator parameter space, then we can conclude that the dark sector in ADM models is either leptophilic, composite, or contains some additional light states into which the dark matter can annihilate (but which does not contribute greatly to the present day energy density).

In this paper, we consider eight possible effective operators linking dark matter with quarks through a weakly coupled UV completion. We ignore some possible additional operators which contain mixed axial/vector or pseudoscalar/scalar interactions ({\it e.g.}~we consider $\bar{\chi}_F\gamma^5 \chi_F \bar{q} \gamma^5 q$ but not $\bar{\chi}_F \chi_F \bar{q} \gamma^5 q$) as the derived bounds are very similar to the ones placed on the operators written below. The operators of interest for complex scalar dark matter (denoted $\chi_S$) are
\begin{eqnarray}
{\cal L}_{S,S} & = & \frac{m_q}{\Lambda^2} \chi_S^* \chi_S \bar{q}q \label{eq:lagSS} \\
{\cal L}_{S,P} & = & \frac{im_q}{\Lambda^2} \chi_S^* \chi_S \bar{q}\gamma^5 q \label{eq:lagSP} \\
{\cal L}_{S,V} & = & \frac{1}{\Lambda^2} \chi_S^*\partial_\mu \chi_S \bar{q} \gamma^\mu q. \label{eq:lagSV}
\end{eqnarray}
Dark matter composed of Dirac fermions is denoted $\chi_F$, and the effective operators under consideration are:
\begin{eqnarray}
{\cal L}_{F,S} & = & \frac{m_q}{\Lambda^3} \bar{\chi}_F \chi_F \bar{q} q \label{eq:lagFS} \\
{\cal L}_{F,P} & = & \frac{m_q}{\Lambda^3} \bar{\chi}_F \gamma^5 \chi_F \bar{q} \gamma^5 q \label{eq:lagFP} \\
{\cal L}_{F,V} & = & \frac{1}{\Lambda^2} \bar{\chi}_F \gamma^\mu \chi_F \bar{q} \gamma_\mu q \label{eq:lagFV} \\
{\cal L}_{F,A} & = & \frac{1}{\Lambda^2} \bar{\chi}_F \gamma^5 \gamma^\mu \chi_F \bar{q} \gamma^5 \gamma_\mu q \label{eq:lagFA} \\
{\cal L}_{F,T} & = & \frac{1}{\Lambda^2} \bar{\chi}_F\sigma^{\mu\nu}\chi_F \bar{q}\sigma_{\mu\nu}q \label{eq:lagFT}
\end{eqnarray}
The second subscript ($S$, $P$, $V$, $A$, or $T$) refers to scalar, pseudoscalar, vector, axial-vector, and tensor interactions respectively, while the first ($S$ or $F$) refer to the spin of the dark matter (scalar or fermion). We have assumed that the coupling to quarks is flavor-blind, and so Eqs.~\eqref{eq:lagSS}-\eqref{eq:lagFT} should be thought of including an implicit sum over all six quark flavors. We shall comment later on the implications of relaxing this constraint. 

Annihilation in the early Universe can proceed through either $s$- or $p$-wave processes (or some combination thereof). The latter case is velocity suppressed, while the former contains terms that are independent of $v$. The interactions in Eqs.~\eqref{eq:lagSV} and \eqref{eq:lagFS} are exclusively $p$-wave. For each operator, we can calculate the cross section times velocity, expanding out to second order in $v$ (see Refs.~\cite{Beltran:2008xg,Fitzpatrick:2010uq} for details):
\begin{eqnarray}
(\sigma |v|)_{S,S} & = & \frac{3}{8\pi\Lambda^4}  \sum_q m_q^2  \left(1-\frac{m_q^2}{m_\chi^2}\right)^{3/2} \label{eq:sigmavSS} \\
(\sigma |v|)_{S,P} & = & \frac{3}{8\pi\Lambda^4} \sum_q m_q^2 \sqrt{1-\frac{m^2_q}{m_\chi^2}}\label{eq:sigmavSP} \\
(\sigma |v|)_{S,V} & = & \frac{3m_\chi^2}{6\pi\Lambda^4} \sum_q \sqrt{1-\frac{m^2_q}{m_\chi^2}} \left(2+\frac{m_q^2}{m_\chi^2}\right)v^2 \label{eq:sigmavSV}
\end{eqnarray}
\begin{eqnarray}
(\sigma |v|)_{F,S} & = & \frac{3m_\chi^2}{8\pi\Lambda^6}  \sum_q m_q^2\left(1-\frac{m_q^2}{m_\chi^2}\right)^{3/2} v^2 \label{eq:sigmavFS} \\
(\sigma |v|)_{F,P} & = & \frac{3m_\chi^2}{2\pi\Lambda^6}  \sum_q m_q^2 \sqrt{1-\frac{m^2_q}{m_\chi^2}} \times\label{eq:sigmavFP}  \\
 & & \left[1+\left(\frac{2m_\chi^2-m_q^2}{8(m_\chi^2-m_q^2)}\right) v^2 \right] \nonumber \\
(\sigma |v|)_{F,V} & = & \frac{3m_\chi^2}{2\pi\Lambda^4}  \sum_q \sqrt{1-\frac{m^2_q}{m_\chi^2}} \times  \label{eq:sigmavFV} \\
 & & \left[ \left(2+\frac{m_q^2}{m_\chi^2}\right)+\left(\frac{8m_\chi^4-4m_q^2m_\chi^2+5m_q^4}{24m_\chi^2(m_\chi^2-m_q^2)}\right)v^2 \right] \nonumber\\
(\sigma |v|)_{F,A} & = & \frac{3m_\chi^2}{2\pi\Lambda^4}  \sum_q \sqrt{1-\frac{m^2_q}{m_\chi^2}} \times \nonumber \\
 & & \left[\frac{m_q^2}{m_\chi^2}+\left(\frac{8m_\chi^4-22m_q^2m_\chi^2+17m_q^4}{24m_\chi^2(m_\chi^2-m_q^2)}\right)v^2 \right]\label{eq:sigmavFA} \\
(\sigma |v|)_{F,T} & = & \frac{3m_\chi^2}{2\pi\Lambda^4}  \sum_q \sqrt{1-\frac{m^2_q}{m_\chi^2}} \times\left[16 \left(1+ \frac{m_q^2}{m_\chi^2}\right)\right. \nonumber \\
 & & \left.+\frac{2}{3}\left(4+\frac{7m_q^2(m_\chi^2+16m_q^2)}{m_\chi^2(m_\chi^2-m_q^2)}\right)v^2 \right]. \label{eq:sigmavFT}
\end{eqnarray}
Effective operators involving leptons rather than quarks would give similar results for $\sigma|v|$, divided by an overall factor of $3$ to account for the quark color.

\begin{figure*}[ht]
\includegraphics[width=0.8\columnwidth]{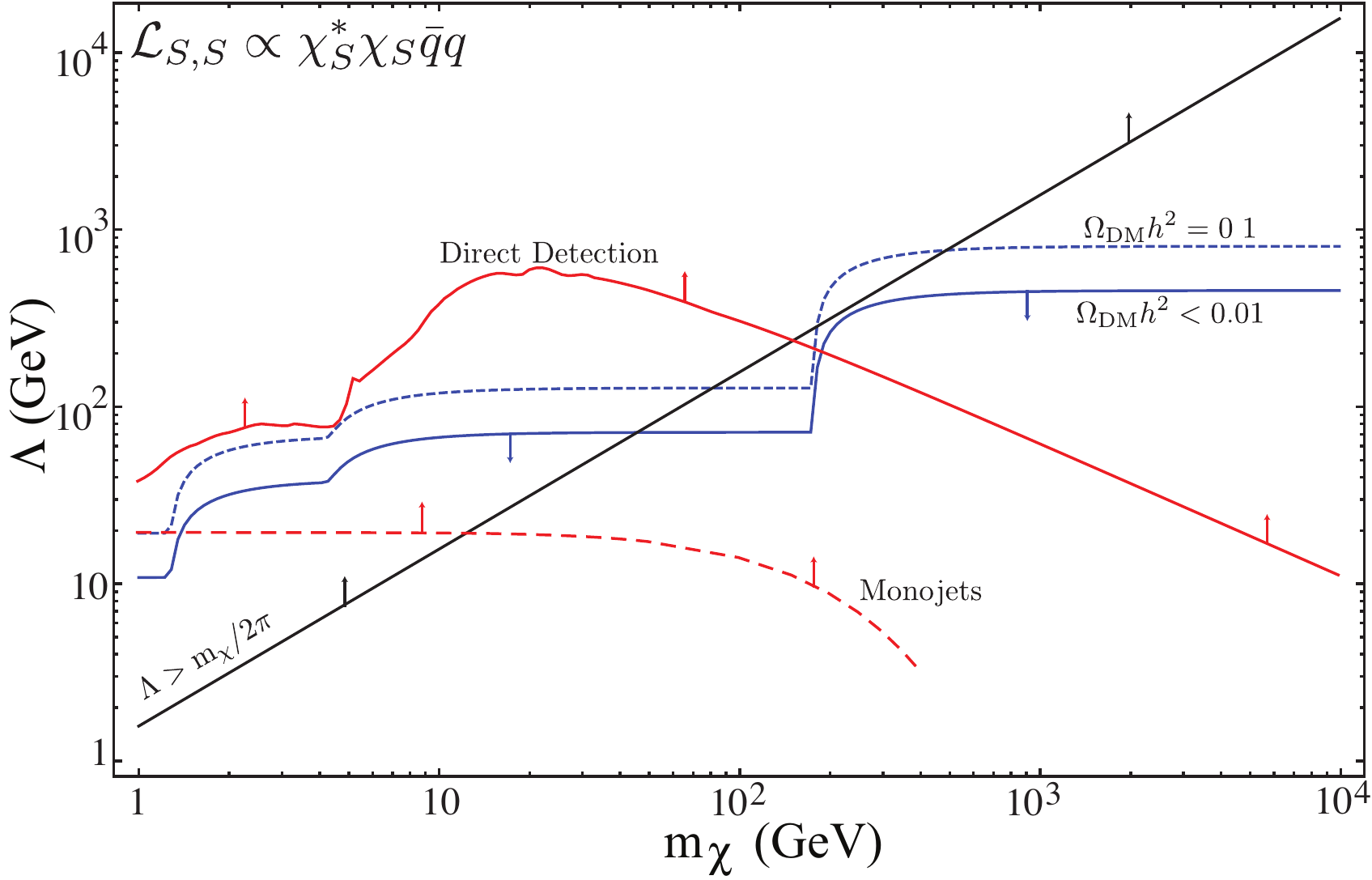}\includegraphics[width=0.8\columnwidth]{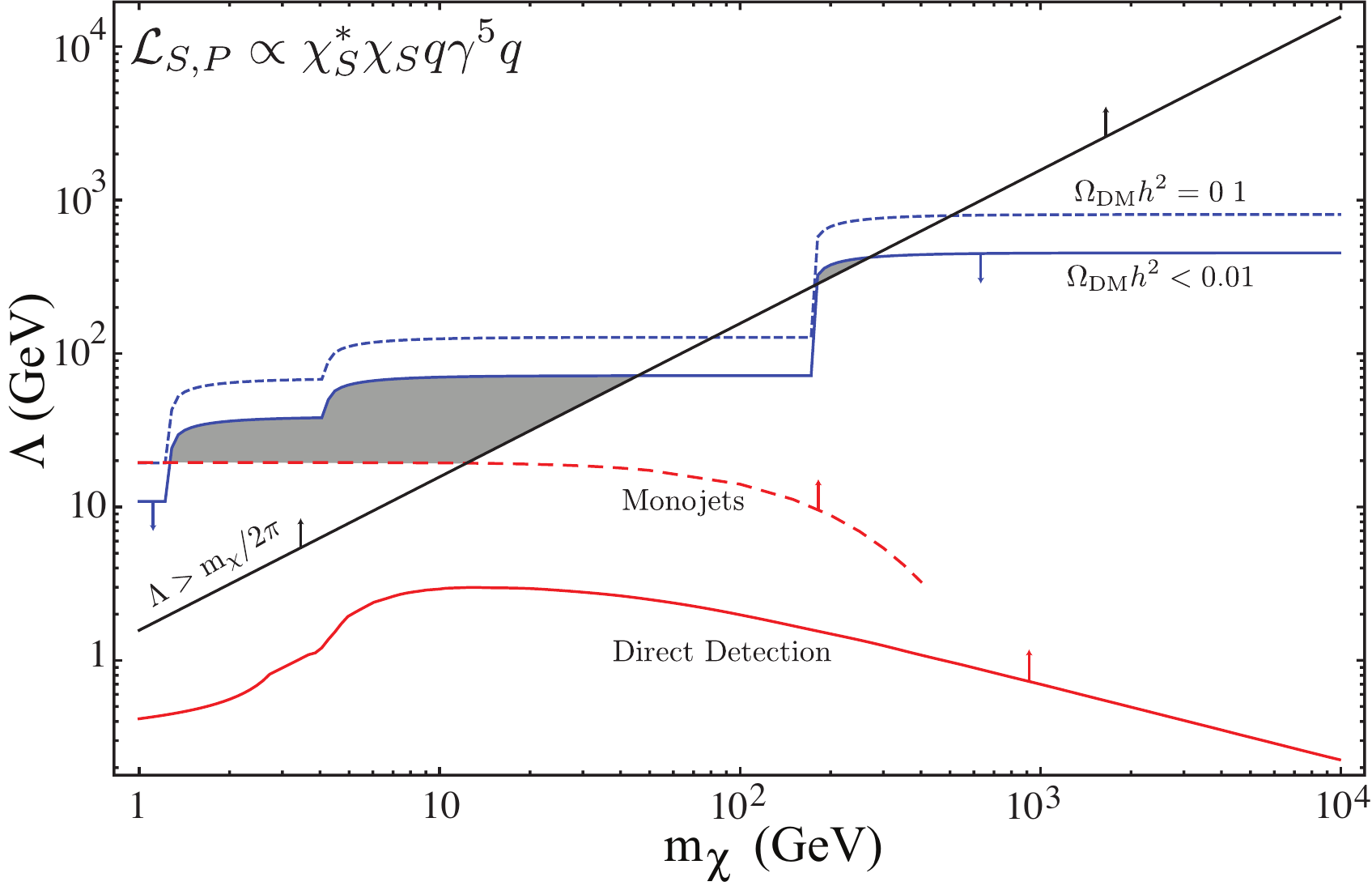}
\includegraphics[width=0.8\columnwidth]{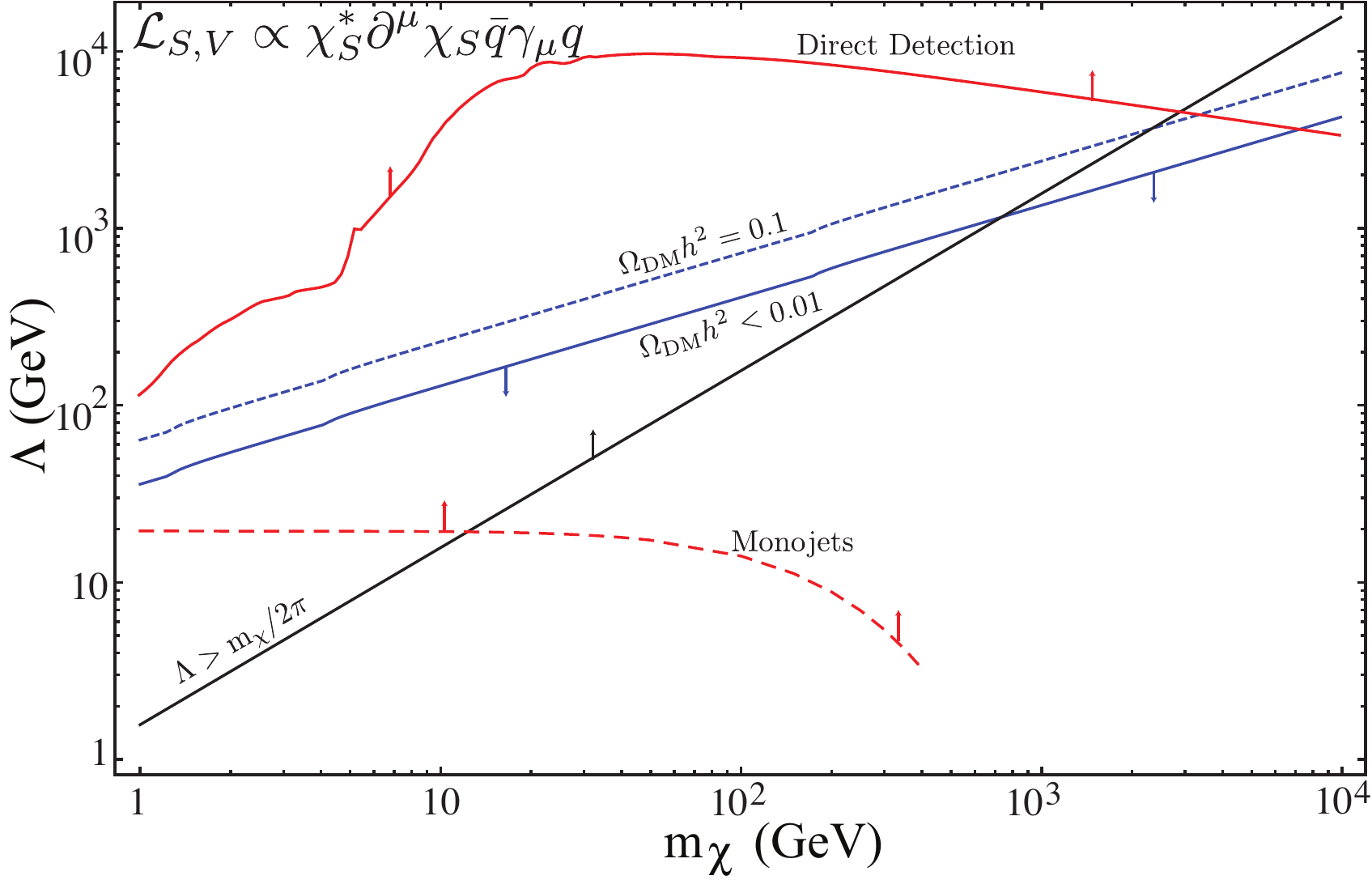}\includegraphics[width=0.8\columnwidth]{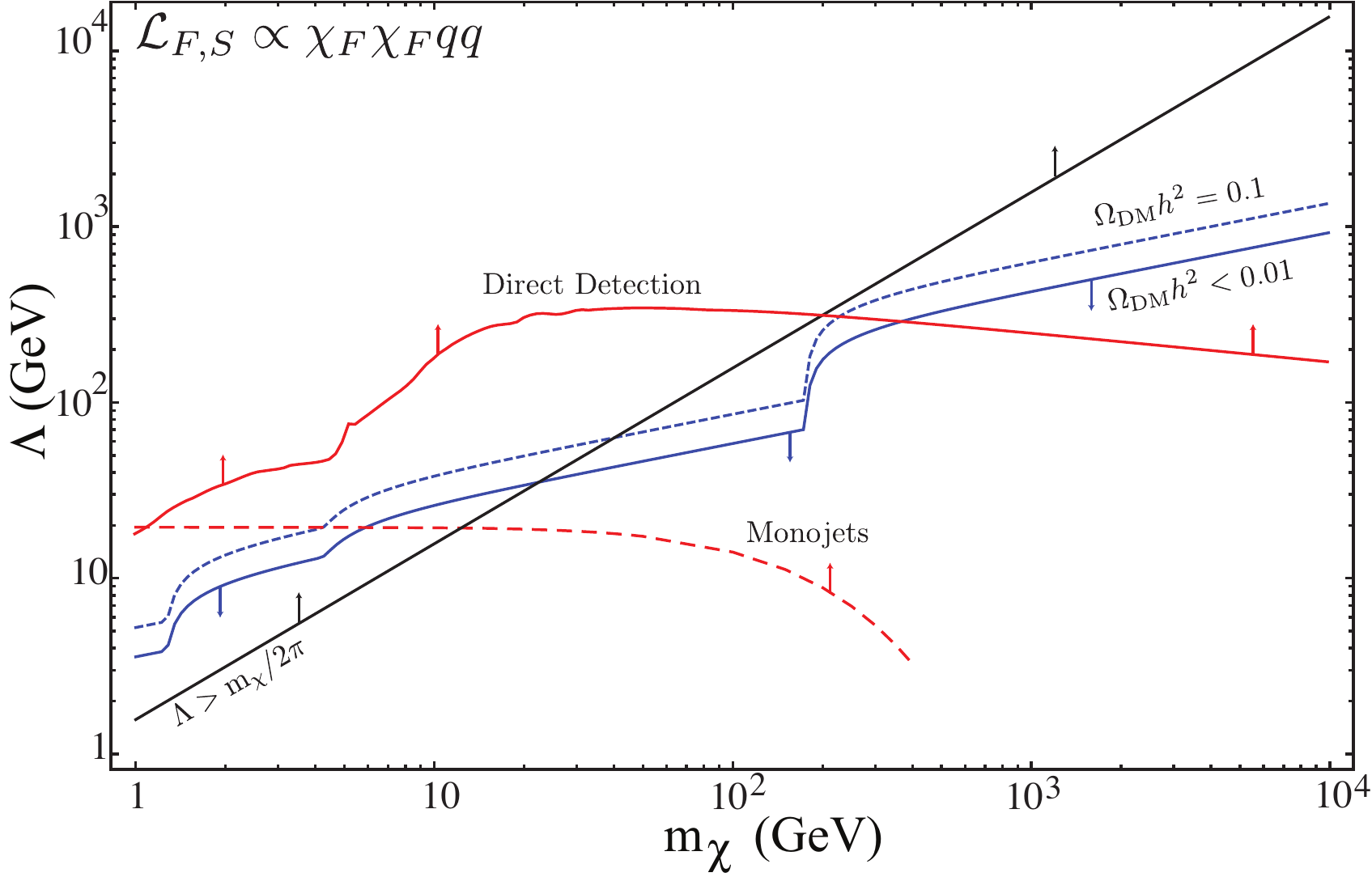}
\includegraphics[width=0.8\columnwidth]{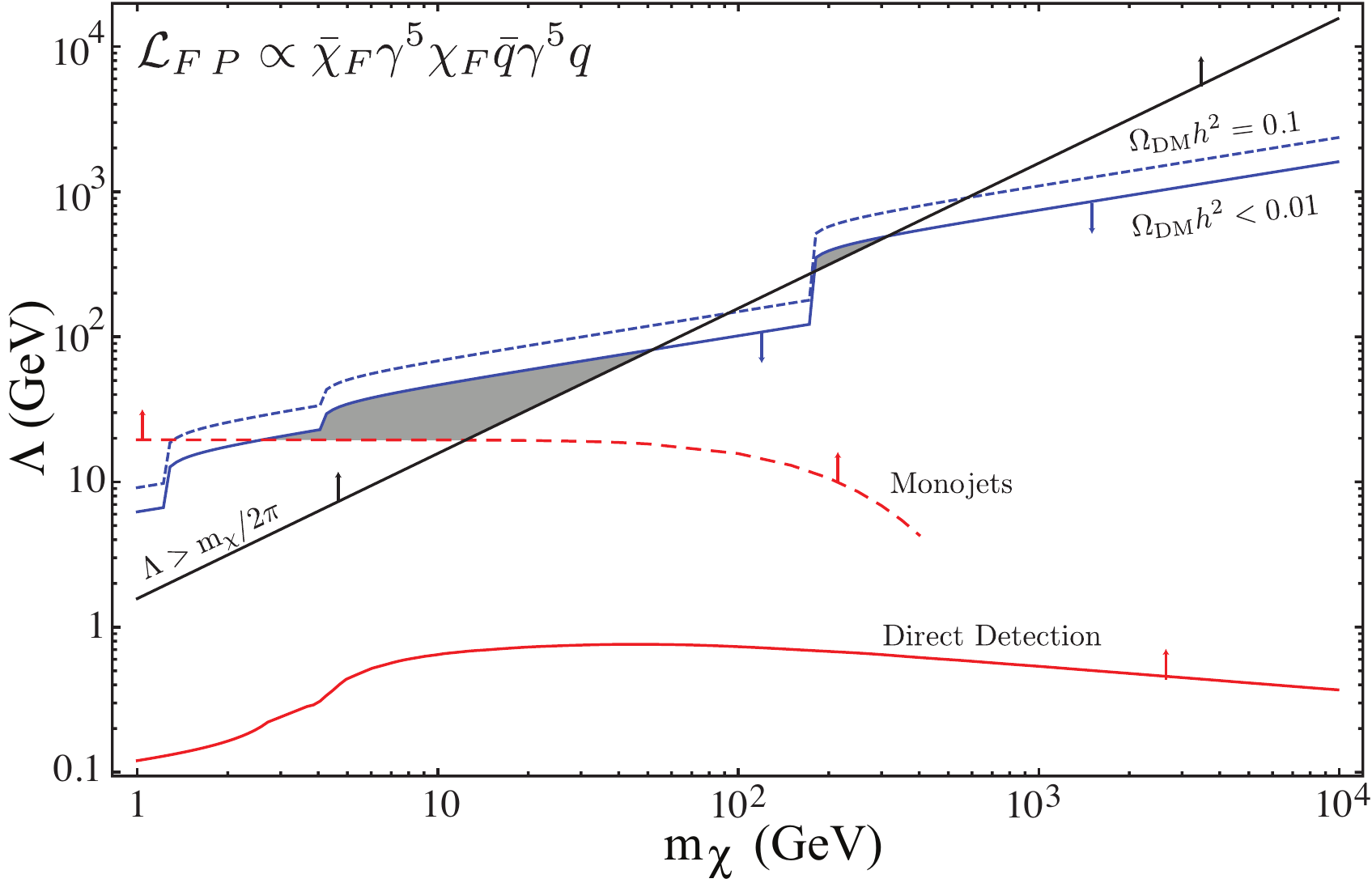}\includegraphics[width=0.8\columnwidth]{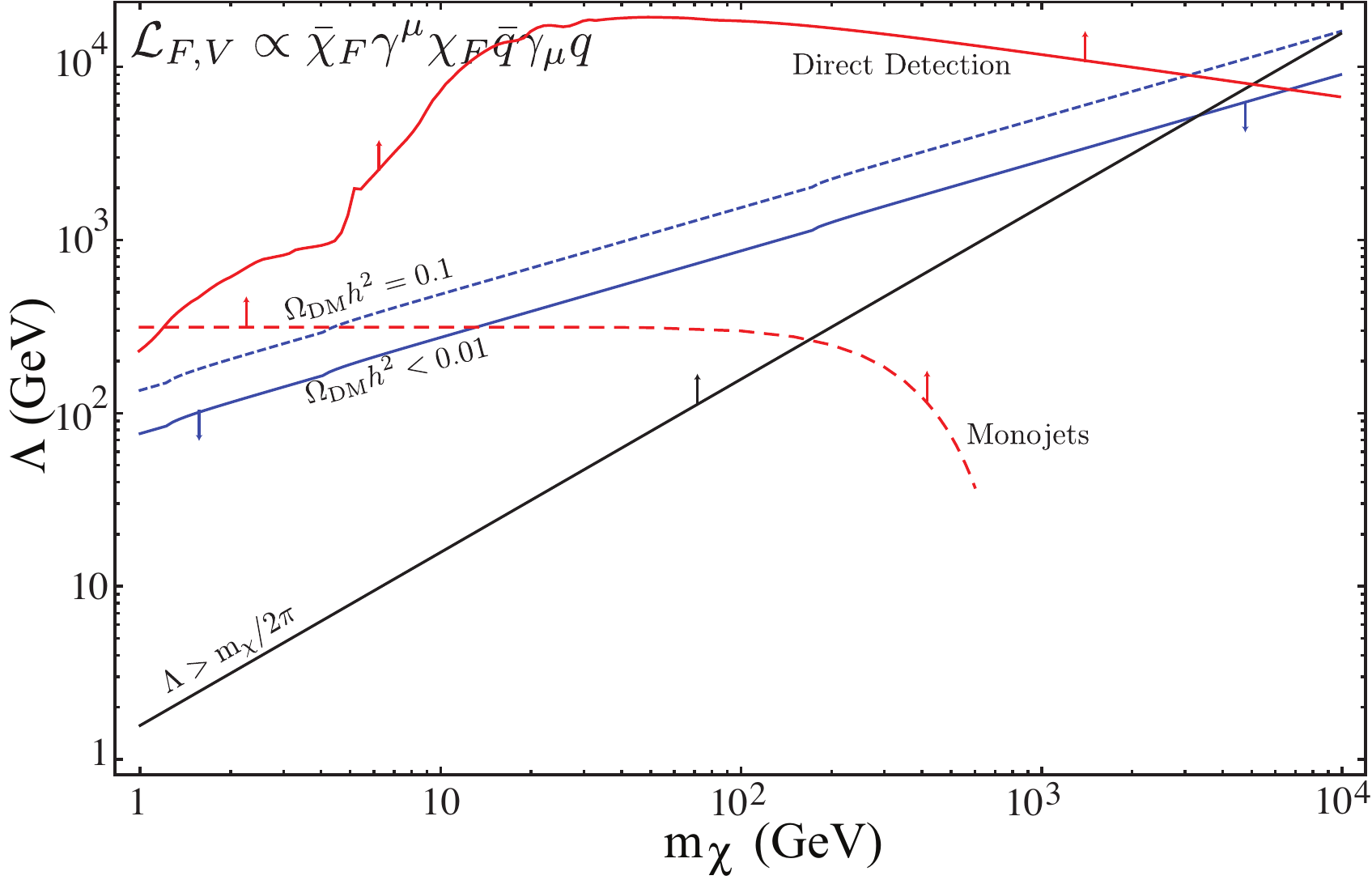}
\includegraphics[width=0.8\columnwidth]{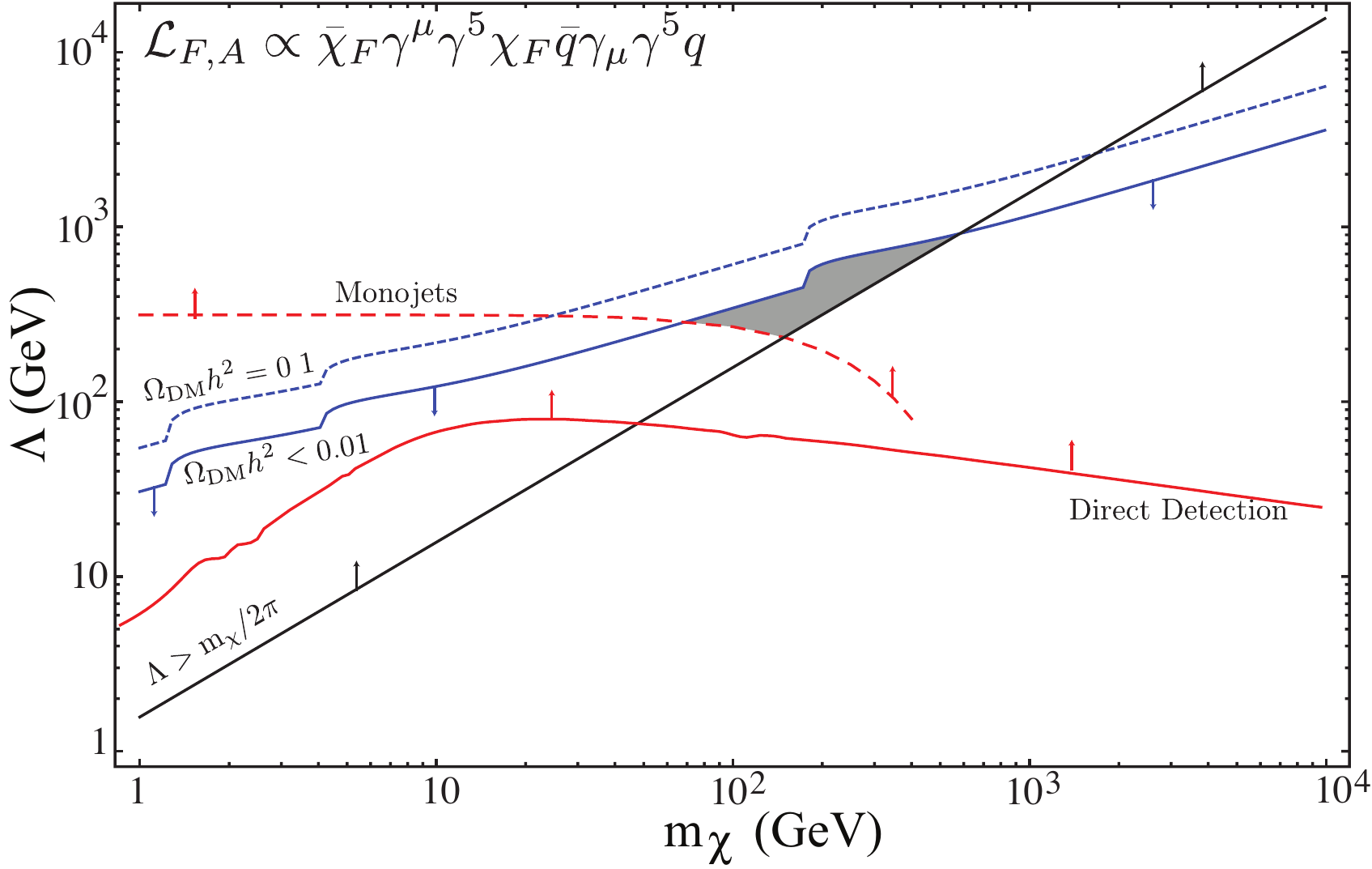}\includegraphics[width=0.8\columnwidth]{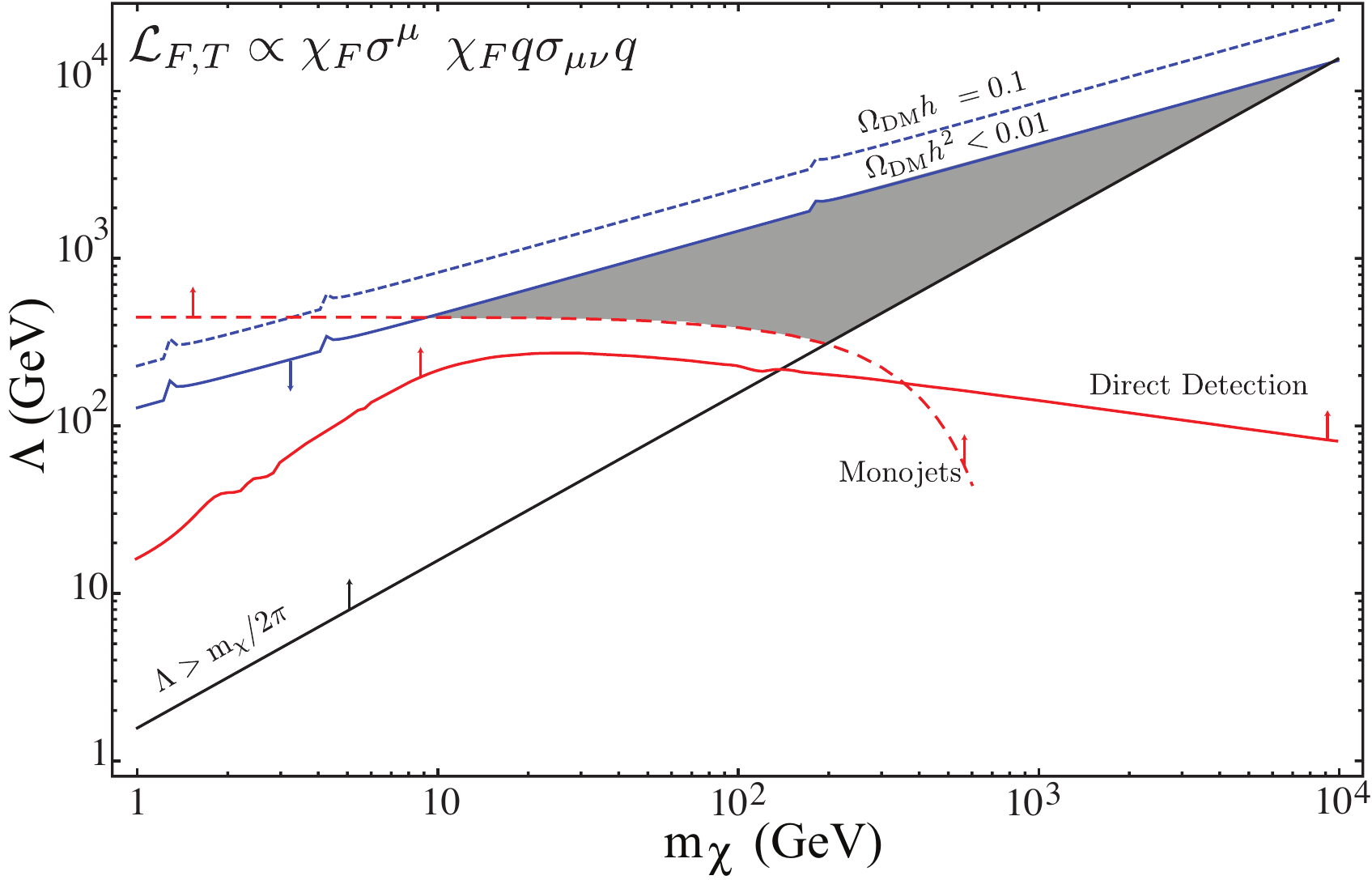}
\caption{Constraints on the scale $\Lambda$ as a function of dark matter mass $m_\chi$ for the eight operators of Eqs.~\eqref{eq:lagSS}-\eqref{eq:lagFT} (in order left to right and descending). Solid blue curve is the upper bound on $\Lambda$ from the requirement that the symmetric component of dark matter compose less than 10\% of the measured value in the Universe (dotted blue is the value of $\Lambda$ that gives the total amount, {\it i.e.}~in a thermal dark matter scenario). Solid red is the lower bound on $\Lambda$ from direct detection experiments. Dashed red is the lower bound on $\Lambda$ from Tevatron monojet searches, taken from Ref.~\cite{Goodman:2010zr} (see also \cite{Bai:2010ys,Goodman:2010ly}). Black solid line shows the lower bound from the requirement that $\Lambda > m_\chi/2\pi$.  Regions above the monojet and direct detection minimum $m_\chi$ which are allowed after all constraints are shown in grey. See text for further details. \label{fig:lambdabounds}}
\end{figure*}

Defining $\sigma |v| \equiv a+bv^2 + {\cal O}(v^3)$, the relic abundance of the symmetric dark matter component after thermal freeze-out is
\begin{equation}
\Omega_{\rm DM} h^2 \approx \frac{(1.04 \times 10^9 ~\mbox{GeV}) x_f}{M_{\rm Pl} \sqrt{g_*} (a+3 b/x_f)}.
\end{equation}
Here, $M_{\rm Pl}$ is the reduced Planck mass, $x_f$ is the ratio of dark matter mass to temperature at freeze-out (detailed calculation shows that $x_f \sim 20-30$ \cite{Kolb:1990vq}), and $\sqrt{g_*}$ is the number of effective degrees of freedom at the time of freeze-out. Requiring that the symmetric dark matter contributes less than $10\%$ of the total, we can place an upper bound on the scale $\Lambda$ of the higher dimensional operators. This choice is somewhat arbitrary, but without significant dilution of symmetric dark matter relative to the asymmetric component, there would be little hope in experimentally differentiating the two (and indeed, little reason to refer to the model as ``asymmetric''). The resulting constraints on $\Lambda$ as a function of $m_\chi$ are shown in Fig.~\ref{fig:lambdabounds}, along with the limits for a thermal WIMP ({\it i.e.}~dark matter who's symmetric component makes up $100\%$ of the dark matter in the Universe). These latter limits are equivalent to those of Ref.~\cite{Beltran:2008xg}. 

Even before considering bounds on the couplings from direct detection, we can already place significant constraints on the  scale $\Lambda$ by requiring that the effective operators arise from a weakly coupled UV completion. In that case, we require that any exchanged particle must have a mass greater than $2m_\chi$. With the additional requirement of perturbative couplings, we find that $m_\chi < 2\pi \Lambda$. As can be seen in Fig.~\ref{fig:lambdabounds}, this requirement severely limits the range of $\Lambda$ and $m_\chi$ that can provide sufficient annihilation for many of the operators, and effectively places an upper bound on the mass of dark matter in these scenarios. However, this bound is somewhat porous, as the factor of $2\pi$ is not a hard limit. Other ${\cal O}(1)$ factors may reasonably be adopted, however this does not qualitatively change our conclusions. While one may certainly imagine non-perturbatively coupled dark matter scenarios ({\it e.g.}~technicolor or composite dark matter \cite{Nussinov:1985xr,Chivukula:1989qb,Bagnasco:vn,Khlopov:2008ly,Frandsen:2011kx,Gudnason:zr}), in those cases it is not possible to calculate the relevant cross sections, and so to make quantitative predictions we must insist that $\Lambda \gtrsim m_\chi /2 \pi$. In any event, a strongly-coupled theory would contain additional states, which as we have noted, are a possible method of evading the bounds derived in this paper.

We next consider the constraints on $\Lambda$ from direct detection. For each operator in Eqs.~\eqref{eq:lagSS}-\eqref{eq:lagFT}, we calculate the resulting spin-dependent or spin-independent elastic scattering cross section as a function of dark matter mass and scale $\Lambda$ \cite{Beltran:2008xg,Fitzpatrick:2010uq,G.Belanger:2008fk}. Comparison to the experimental upper limits on the nucleon-DM scattering cross section $\sigma_{\chi N}$ allows us to place lower limits on $\Lambda$ as a function of $m_\chi$. Note that for $m_\chi \lesssim 5$~GeV, no bounds are set by the current experiments.

The strength of the direct detection bounds depends greatly on whether the dark matter interacts with nucleons via spin-dependent or spin-independent interactions. Of the effective operators of interest in this paper, the scalar and vector interactions (Eqs.~\eqref{eq:lagSS}, \eqref{eq:lagSV}, \eqref{eq:lagFS}, and \eqref{eq:lagFV}) induce spin-independent scattering, while the fermionic axial and tensor interactions (Eqs.~\eqref{eq:lagFA} and \eqref{eq:lagFT}) result in spin-dependent scattering \cite{Jungman:1995df}. Note that the pseudoscalar interactions (Eq.~\eqref{eq:lagSP} and \eqref{eq:lagFP}) do not lead to either spin-dependent or spin-independent couplings that are velocity independent. We include the derived bounds from the resulting spin-dependent direct detection cross section \cite{Bai:2010ys,Cheng:1988im}, which are proportional to powers of the momentum transfer $q = \sqrt{2m_\chi E_R}$ (here $E_R$ is the energy of the recoiling nucleon; we assume $E_R \sim 50$~keV). As can be seen in Fig.~\ref{fig:lambdabounds}, the resulting bound on $\Lambda$ from these $q$-dependent interactions are extremely weak; in fact, they require a mediator mass typically less than the mass of the dark matter. 

The spin-independent constraints are taken from CDMS \cite{Kamaev:2009vn,Collaboration:2010nx}, CoGeNT \cite{Aalseth:2008rx}, CRESST \cite{Altmann:fk}, XENON-10 \cite{Angle:2007uj}, and XENON-100 \cite{Aprile:2010um}. For dark matter with mass between $\sim 5- 80$~GeV, XENON-10 and XENON-100 provide the best limits at the present time. Above this range, CDMS has the best constraint. For very low masses, near $\sim 1$~GeV, the CRESST detector has the most stringent bounds. In the intermediate region, CoGeNT and the CDMS low threshold  \cite{Collaboration:2010nx} dominate. Spin-dependent constraints are a combination of COUPP \cite{Behnke:2008kl}, CRESST \cite{Altmann:fk}, and PICASSO \cite{Archambault:2009oq}, the last of these providing the best limits between $\sim 5-100$~GeV, while CRESST dominates in the very low mass window.

The underlying assumption should again be noted: we are requiring that the same operators responsible for the thermal relic abundance will be responsible for any direct detection interaction. If, for example, the dark matter had large couplings to leptons ({\it i.e.}~small $\Lambda$ for $\bar{\chi}\chi \ell \bar{\ell}$-type operators), this could provide sufficient suppression of the symmetric component, while a separate operator would be responsible for direct detection. We will have more to say on this in the conclusion.

Finally, we can consider constraints on effective operators searches at the Tevatron \cite{Bai:2010ys,Goodman:2010ly,Goodman:2010zr}. We use the bounds from Ref.~\cite{Goodman:2010zr}, as all the operators in Eqs.~\eqref{eq:lagSS}-\eqref{eq:lagFT} are considered. For each operator, the resulting cross section for dark matter pair production plus an extra jet was compared to the searches performed in the monojet $+\slashed{E}_T$ channel at CDF \cite{CDFmonojet}. As the experimental data is in agreement with the Standard Model, lower bounds are placed on $\Lambda$ for each operator. While the range of dark matter masses that can be probed by this method is limited, these constraints do provide a bound which is in several cases complimentary to that provided by direct detection.

We show the remaining allowed regions in Fig.~\ref{fig:lambdabounds} in grey. In total, we see that the four requirements (over-annihilation, direct detection, Tevatron monojets, and consistency of the effective operator expansion), completely exclude four classes of asymmetric dark matter over the entire mass range. It is interesting to note that this is also true for many effective operator expansions for WIMP thermal dark matter, as has been noted before \cite{Beltran:2008xg}, and can be seen in Fig.~\ref{fig:lambdabounds} by considering the $\Omega_{\rm DM}h^2 =0.1$ line for $\Lambda$.

Several windows in $\Lambda$ vs.~$m_\chi$ space remain open for pseudoscalar, axial, and tensor operators. Many of these are in the low mass ($\sim5-10$~GeV) region, which is especially intriguing in the context of ADM. Monojet searches at the LHC with 100~fb$^{-1}$ of data may reasonably be expected yield a factor of ${\cal O}(10)$ improvement over the Tevatron bounds and extend them to higher $m_\chi$ \cite{Goodman:2010zr}. This would allow most of the remaining pseudoscalar and axial windows to be closed, and greatly reduce the allowed parameters for the tensor case. It is clear then, that asymmetric dark matter coupled to quarks via effective operators are highly constrained by the data.

We now briefly mention the impact of generation or flavor dependent couplings on the bounds for $\Lambda$. Clearly, we cannot perform an exhaustive analysis, as there are a limitless set of possible flavor-dependencies that can be added to the effective operators. In general, coupling to fewer generations will make the upper bounds coming from the over-annihilation constraint much more restrictive. As there are fewer Standard Model particles involved in the annihilation process, each particle which does interact must do so more efficiently. Thus, $\Lambda$ must be lower. Combined with the $\Lambda > m_\chi/2\pi$ bound, this can completely exclude scalar and pseudoscalar mediators, for example if the dark matter couples only to $u$ and $d$ quarks.

The changes to direct detection bounds in a flavor-dependent scenario are more complicated. Vector mediators depend only on the couplings to $u$ and $d$ quarks, so restricting couplings only to these two flavors will not change the bounds, while coupling to the heavier generations will completely eliminate it. Scalar, axial, and tensor interactions depend most heavily on $u$, $d$, and $s$ couplings; eliminating these will cause the bounds to loosen considerably. 

Flavor dependent constraints from monojets are investigated in Ref.~\cite{Bai:2010ys}. Couplings to $u$ quarks are the most constrained, followed by $d$ and $s$, as expected for results from a proton-antiproton collider. Reducing the number of quarks that couple to dark matter allows each coupling to be larger, thus setting a less restrictive lower limit on $\Lambda$, as can be see by comparison of the results of Ref.~\cite{Bai:2010ys} and Ref.~\cite{Goodman:2010zr}.

Returning to the general case, even when certain operators are completely excluded, it is obvious that many possibilities exist to which would allow us to escape the conclusions of the analysis presented above. For example, the annihilation could not proceed through dark matter-quark interactions, or perhaps the assumption that the annihilation proceeds through an effective operator could be incorrect. In both cases, this communicates valuable information about the structure of any asymmetric dark matter model. Let us consider each in turn.

If we imagine asymmetric dark matter avoids the constraints derived in this paper by annihilating primarily into some other light field, then one possible explanation is that the fundamental field which was integrated out in the effective operator is leptophilic. This is an intriguing possibility in light of the models of leptophilic dark matter (see, for example, Ref.~\cite{Arkani-Hamed:2008vn,Cholis:2008kx,Cholis:2008uq,Fox:2008fk,Essig:2009ys,Kohri:2009ys}) which attempt to explain anomalies in the PAMELA positron fraction \cite{Adriani:2008bh} and Fermi Gamma-Ray Space Telescope $e^++e^-$ spectrum \cite{Collaboration:2009dq}. Effective operators involving leptons would not be greatly constrained by direct detection, however the over-annihilation requirement would remain, as would the $\Lambda > m_\chi/2\pi$ constraint. The monojet search could be replaced by a monophoton search at LEP \cite{Fox:2011tg}, though the mass range would be limited.

Alternatively, the dark matter could annihilate efficiently into some new dark state that is either very light or unstable, decaying into Standard Model particles before Big Bang Nucleosynthesis (BBN) (see for example Ref.~\cite{Hall:2010uq}). In the former case, CMB and BBN constraints on the number of relativistic species (usually stated in terms of the number of neutrino flavors) must be avoided. This could be achieved through significant entropy injection into the thermal bath after dark matter annihilation decouples \cite{Ackerman:2008zr}. In any event, this possibility requires an extended dark sector in addition to the dark matter and the high-scale mediator.

Finally, the results of this paper could be interpreted to mean that the annihilation and scattering of asymmetric dark matter cannot be written in terms of an effective field theory. This may mean that the coupling to the mediator is non-perturbative; as in the case of quirky asymmetric dark matter \cite{Kribs:2009ve} or composite \cite{Nussinov:1985xr,Chivukula:1989qb,Bagnasco:vn,Khlopov:2008ly,Frandsen:2011kx,Gudnason:zr}. Alternatively, the mediator mass could simply be lower than the cutoff of $m_\chi/2\pi$. This is an interesting possibility, because it requires that ADM not be ``maverick'' \cite{Beltran:2010cr}. That is, additional light states would be required in order to satisfy all the experimental constraints. Should this possibility be born out, this would again be very interesting in the context of the light mediator solutions \cite{Arkani-Hamed:2008vn,Bjorken:2009mm,Cholis:2008qq,Cholis:2008vb,Cholis:2008wq,Hisano:2003ec,Pospelov:2008jd} of the PAMELA and Fermi anomalies.

In this paper, we have investigated effective operators suppressed by a scale $\Lambda$ connecting dark matter to quarks in the context of asymmetric dark matter models. The large couplings (compared to that of a WIMP model) required by the over-annihilation of ADM in the early Universe, combined with experimental constraints from direct detection and Tevatron searches greatly constrain the parameter space of $\Lambda$ as a function of dark matter mass $m_\chi$. Only relatively narrow windows -- including several at low mass where many ADM models prefer the dark matter to be -- remain for most of the operators, and future LHC data can be expected to close many of these.

As efficient thermal annihilation is the one universal requirement of ADM, we consider it useful to clearly set forth the bounds which any such model must satisfy. Most currently existing models of ADM evade these constraints though various method. In this paper, we have outlined several general techniques for doing so:
\begin{itemize} 
\item by annihilation into leptons,
\item by annihilation into additional dark states which are very light or unstable,
\item low mass mediators that cannot be written as effective operators,
\item new confining gauge groups in the dark sector.
\end{itemize} 
It is interesting to consider the implications of these possibilities. Outside a narrow range of parameters, the experimental constraints seem to push asymmetric dark matter into scenarios which are either leptophilic or contain additional light states.

\section*{Acknowledgements} 

The author thanks Scott Dodelson, Graham Kribs, Roni Harnik, Dan Hooper, Patrick Fox, Hugh Lippincott, Ethan Neil, Will Shepherd, Ian Shoemaker, and Tim Tait for their advice and suggestions.

\bibliography{effectiveops}
\bibliographystyle{apsrev}

\end{document}